\documentclass[aps,prc,longbibliography,twocolumn,nofootinbib]{revtex4-2}
\usepackage{physics}
\usepackage{amsmath}
\usepackage{amssymb}
\usepackage{graphicx}
\usepackage{float}
\usepackage{array}
\usepackage{verbatim}
\usepackage[margin=1in]{geometry}
\usepackage{color}
\usepackage{natbib}
\usepackage{subcaption}
\graphicspath{{images/}}

\usepackage{dcolumn}
\usepackage{bm}
\usepackage{physics}
\usepackage{hyperref}

\usepackage[list=true]{subcaption}
\usepackage{placeins}

\usepackage{enumitem}
\usepackage{calc}

\begin{document}

\title{Shell effects in quasi-fission in reactions forming $^{226}$Th compound nucleus}
\author{H. Lee}
\author{P. McGlynn}
\author{C. Simenel}
\affiliation{Department of Fundamental and Theoretical Physics and Department of Nuclear Physics and Accelerator Applications, Research School of Physics,
The Australian National University, Canberra ACT 2601, Australia}
\date{\today}

\begin{abstract}
\begin{description}
\item[Background]{
Quasi-fission reactions occur in fully damped heavy-ion collisions without the formation of an equilibrated compound nucleus, leading to the formation of fragments with similar properties as in fission reactions. 
In particular, similar shell effects are expected to affect fragment formation in both fission and quasi-fission.
Experimentally, the role of shell effects in quasi-fission is still debated and further theoretical predictions are needed. 
}
\item[Purpose]{
To investigate quasi-fission dynamics in different reactions forming the same compound nucleus and search for possible signatures of shell effects in fragment formation. 
}
\item[Methods]{$^{50}$Ca$+^{176}$Yb and $^{96}$Zr$+^{130}$Sn  quasi-fission reactions are simulated with the time-dependent Hartree-Fock code Sky3D near the Coulomb barrier. Evolutions of the quadrupole ($Q_{20}$) and octupole ($Q_{30}$) moments are interpreted in terms of features of the potential energy surface (PES) of the $^{226}$Th compound nucleus.}
\item[Results]{
Both reactions encounter quasi-fission. In $^{50}$Ca$+^{176}$Yb, those only occur at finite angular momenta. In the more symmetric  $^{96}$Zr$+^{130}$Sn reaction with stronger Coulomb repulsion in the entrance channel, quasi-fission also occurs in central collisions. In agreement with earlier predictions, $^{50}$Ca$+^{176}$Yb encounters partial mass equilibration that is stopped when the heavy fragment reaches $Z\approx54$ protons, as in the asymmetric fission mode of $^{226}$Th. Interestingly, $^{96}$Zr$+^{130}$Sn encounters an ``inverse quasi-fission'' (multi-nucleon transfer increasing the mass asymmetry between the fragments) also leading to similar fragments as in asymmetric fission. In both systems, quasi-fission trajectories in the $(Q_{20}-Q_{30})$ plane are found close to the asymmetric fission valley of $^{226}$Th PES. 
}
\item[Conclusions]{
The observation of an inverse quasi-fission that goes against expectations from a simple liquid drop picture suggests that shell effects have an influence in quasi-fission. In addition, the similarity between fragments formed in asymmetric fission and quasi-fission  supports the idea that the same shell effects are at play in both mechanisms. 
In particular, these were recently attributed to octupole deformed shell effects in $Z=52-56$ fragments. 
Interpreting quasi-fission dynamics with PES used in fission is naturally limited by the fact that these PES are usually computed with axial symmetry, no angular momentum and no excitation energy, thus motivating future developments of PES for quasi-fission. }
\end{description}
\end{abstract}
\maketitle

\section{Introduction}

Nuclear fission is one of the most complex nuclear processes, thus challenging theoretical many-body modelling~\cite{bender2020,schunck2022}. 
In particular, accounting for quantum shell effects in the fissioning system \cite{gustafsson1971,ichikawa2019,cwiok1994,bernard2023} as well as in the fragments \cite{mayer1948,meitner1950,wilkins1976,scamps2018,scamps2019,bernard2023} is necessary to explain fission properties such as fragment mass asymmetries observed in experiments \cite{unik1974,schmidt2000,bockstiegel2008,chatillon2019,mahata2022,schmitt2021b,chatillon2022}. 
Shell effects may also influence the role of dissipation in fission as well as fission time \cite{ramos2023}.

Shell effects are largely responsible for  the topography of potential energy surfaces (PES) that represent the minimum energy of the system under a set of constraints on its shape, such as quadrupole and octupole moments that are traditionally used to constrain the elongation and asymmetry of the system. 
In particular, shell effects are able to induce fission valleys driving the system towards asymmetric fission. 
In fact, shell correction energy \cite{strutinsky1967,strutinsky1968} and single particle level density near the Fermi level \cite{bernard2023} show that several shell effects are at play until scission. 
Then, the final asymmetry is influenced by shell effects in the pre-fragments, such as octupole deformed shell effects in the heavy fragment of actinide fission \cite{scamps2018}. 

Shell effects are also expected to play a role in quasi-fission reactions. 
The latter occur in fully damped heavy-ion collisions with mass transfer between the fragments, usually producing outgoing fragments that are  more symmetric  than in the entrance channel \cite{toke1985} (see \cite{hinde2021} for a recent review on experimental studies of quasi-fission). 
Typical contact times between the fragments in quasi-fission are of the order of few $10^{-20}$~s \cite{toke1985,durietz2013}, i.e., similar to mass equilibration time scale but much slower than other equilibration and dissipation processes   \cite{williams2018,simenel2020}.
Although similar fragments can be produced in quasi-fission and in fusion followed by fission, there is no formation of an equilibrated compound nucleus in quasi-fission and its outgoing fragments could keep a ``memory'' of the entrance channel. 
In fact, quasi-fission is the main mechanism that hinders the formation of superheavy compound nuclei in fusion reactions.

Mass-equilibration is expected to stop when shell effects are present in the fragments that prevent further transfer of nucleons. 
For instance, the formation of fragments in the vicinity of $^{208}$Pb in reactions with actinide targets is expected to be favoured by its spherical shell effects. 
Experimental signatures have been discussed in the literature \cite{itkis2004,nishio2008,kozulin2014,wakhle2014,morjean2017,hinde2018}.
However,  it has been recently proposed that sequential fission of the target like fragment could be responsible for the  peak at $A\approx208$ nucleons obtained after rejecting events with three nuclei in the exit channel \cite{jeung2022}. 
Reactions with sub-lead targets should be free of sequential fission and thus avoid this difficulty. In addition, reactions forming actinide compound nuclei can be used to compare quasi-fission products with  fission modes that are usually known in this region. 
Nevertheless, even for such lighter systems the influence of shell effects in quasifission is debated, with some experiments finding that  ``shell effects are clearly seen'' \cite{chizhov2003}, while others only conclude of ``weak evidences'' \cite{hinde2022}. 

Several theoretical approaches have been used to investigate quasi-fission mechanisms, including 
the dinuclear system model \cite{adamian2003,huang2010,guo2018c},  models based on the Langevin equation~\cite{zagrebaev2005,aritomo2012,karpov2017}, 
molecular dynamics~\cite{wang2015,wang2016,klimo2019,li2020b}, the Boltzmann-Uehling-Uhlenbeck model \cite{feng2023}, and the
time-dependent Hartree-Fock (TDHF) theory and its extensions \cite{wakhle2014,oberacker2014,hammerton2015,umar2015a,umar2016,sekizawa2016,yu2017,ayik2018,morjean2017,godbey2019,simenel2021,li2022,mcglynn2023}
(see~\cite{simenel2012,simenel2018,sekizawa2019,stevenson2019,godbey2020} for reviews on TDHF).

Recent TDHF simulations of quasifission in   $^{50}$Ca$+^{176}$Yb collisions predicted that the mass equilibration process stops when the heavy fragment reaches $Z\approx54$ protons \cite{simenel2021}. 
The corresponding mass and charge asymmetries, as well as the total kinetic energy (TKE) of the fragments were shown to match those of the fragments produced in the asymmetric fission mode of the $^{226}$Th compound nucleus.
These observations support the fact that similar shell effects affect both fission and quasifission. 
The reason for choosing this system was that $^{226}$Th exhibits both a symmetric and an asymmetric fission modes. 
However, the symmetric mode was not observed in $^{50}$Ca$+^{176}$Yb quasifission, which led to the conclusion that not all fission modes are expected to be necessarily produced in quasi-fission.
Several systems leading to the same $^{294}$Og compound nucleus were also studied with TDHF in a more recent work~\cite{mcglynn2023}. 
It was shown that at least some quasifission reactions could be interpreted in terms of the topography of the PES despite the simplifications used to compute the latter (no excitation energy, zero angular momentum, and axial symmetry).
A similar approach is adopted in the present work where the topography of the $^{226}$Th PES is used to interpret quasifission dynamics in $^{50}$Ca+$^{176}$Yb and $^{96}$Zr$+^{130}$Sn reactions. 
The second reaction, being more symmetric than the asymmetric fission mode of $^{226}$Th, was chosen with the anticipation that more symmetric quasifission fragments could be produced if they were driven by the symmetric fission valley. 

The details of PES and TDHF calculations are described in section \ref{sec:Method}.
The quasifission simulations with TDHF are analysed in section \ref{sec:results}. 
The quasifission dynamics is interpreted in terms of the $^{226}$Th PES in section \ref{sec:compare}.
Conclusions are drawn in section \ref{sec:conc}.
Tables summarising TDHF results are provided in Appendix.

\section{Method\label{sec:Method}}

\subsection{PES and fission modes}
The PES is a landscape of nuclear potential energies associated with mean-field states of various nuclear shapes. 
It is commonly computed in terms of multipole moments. In particular the quadrupole moment 
\begin{equation}
    Q_{20} = \sqrt{\frac{5}{16\pi}}\int \dd[3]{r} \rho(\textbf{r})(2z^2-x^2-y^2),
\end{equation}
provides a proxy for the elongation of the system while the octupole moment
\begin{equation}
    Q_{30} = \sqrt{\frac{7}{16\pi}}\int \dd[3]{r} \rho(\textbf{r})(2z^3-3z(x^2+y^2))
\end{equation}
is a measure of its asymmetry. 

\begin{figure}[!htb]
    \centering
    \includegraphics[width=\linewidth]{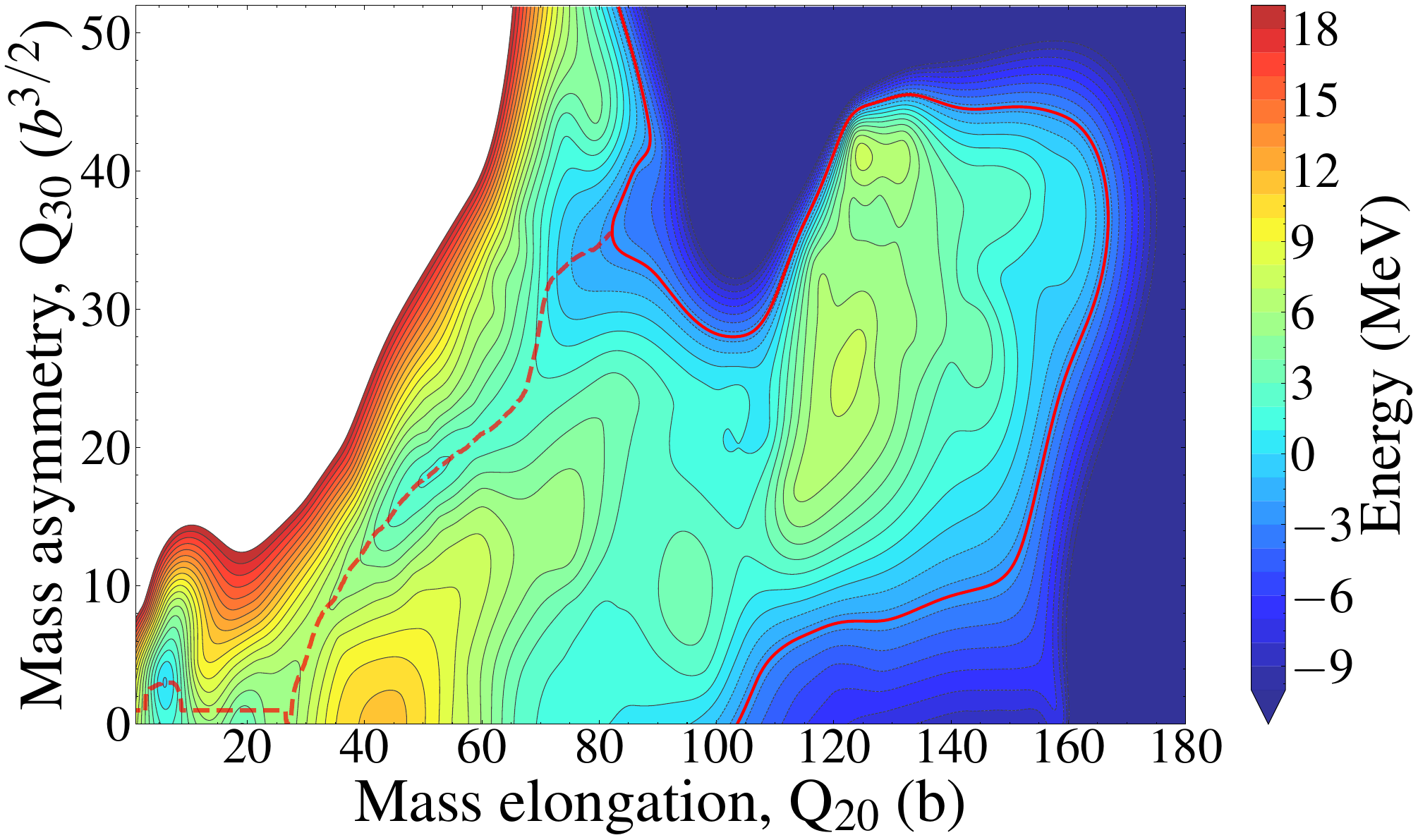}
    \caption{PES of $^{226}$Th obtained using SkyAx with quadrupole and octupole steps $\Delta q_{20} =1.44$~b and $\Delta q_{30} =1$~b$^{3/2}$, respectively.
The dashed line shows the minimum energy fission path. The solid red line is the scission line. }
    \label{fig:PES}
\end{figure}

Here, the PES is computed from the constrained Hartree-Fock (HF) theory with axial symmetry on nuclear shapes.
The states used to build the PES have no internal excitation and their average  angular momentum is zero. 
The PES of $^{226}$Th shown in Fig.~\ref{fig:PES} was calculated with constraints on $Q_{20}$ and $Q_{30}$  with the SLy4$d$ Skyrme functional \cite{kim1997} and  BCS pairing interaction with density dependent delta interaction using the SkyAx code \cite{reinhard2021}.

The minimum energy fission path from the ground-state to scission corresponds to the dashed line in Fig. \ref{fig:PES}.
This is the path taken by increasing $Q_{20}$ and choosing $Q_{30}$ to give the minimal energy.
The valley which the dashed line follows corresponds to the asymmetric fission valley, whereas the symmetric fission valley leads to scission configurations at $Q_{30} \simeq 0$.
The scission line is determined as a set  of points in the PES where the neck density falls below $\rho=0.08$~fm$^{-3}$.

\subsection{TDHF simulations}\label{sec:TDHF}

The TDHF mean-field theory can be obtained from a variational principle on Dirac action, leading to an equation on the one-body density matrix~$\rho$
\begin{equation}
i\hbar \frac{d\rho}{dt}=\left[h[\rho],\rho\right]
\end{equation}
where ${h}[\rho]$ is the Hartree-Fock single-particle Hamiltonian. 
Here, it is obtained from a Skyrme energy density functional $E[\rho]$ according to $h[\rho]_{\alpha\beta}=\frac{\delta E[\rho]}{\delta\rho_{\beta\alpha}}$. 
The HF Hamiltonian is self-consistent as it depends on the one-body density matrix $\rho$ of the system.

The ground-states of the collision partners are first obtained from a static mean-field calculation on a cartesian grid of  28$\times$28$\times$28~fm$^{3}$  with mesh size $\Delta x=1$~fm.
As in PES calculations, the SLy4$d$ Skyrme functional \cite{kim1997} was used with the same pairing functional to account for pairing correlations at the BCS level. 
The resulting ground-state wave-functions of collision partners are then placed in a larger cartesian grid with the same mesh size, with an initial distance of 56~fm between the centers of mass. 
The collision axis between the fragments were aligned with the z-axis of the simulation, with the mid-point between the centers of mass of two fragments placed at the origin. 
Shape coexistence is expected in $^{96}$Zr with $\beta_2\simeq 0.13$ prolate ($\beta_2\simeq-0.14$ oblate) energy minimum only 0.25~MeV (0.13~MeV) below the spherical configuration. 
It was then constrained to be spherical to probe the average behaviour between prolate and oblate deformations.
Furthermore, the spherical $^{96}$Zr was time-evolved alone in a simulation box to ensure that fluctuations in shape would be minimal over the timescale of the collision.
Negligible changes in the quadrupole and octupole moments were found during the time evolution indicating stability of the spherical mean-field configuration. 
The $^{176}$Yb nucleus is found with a significant prolate deformation so to consider effects of orientation of the target nucleus with respect to the  collision axis, the $^{176}$Yb nucleus was prepared in two configurations in a grid, rotated by $\frac{\pi}{2}$ radians.
For non-central collisions of $^{50}$Ca$+^{176}$Yb, the side and tip orientations refer to the initial orientation of $^{176}$Yb with respect to the z-axis of the simulation.

A Galilean boost is  applied on each nucleus according to the required centre of mass energy $E_{c.m.}$ and the initial orbital angular momentum $L$, assuming that prior to this initial condition, the nuclei followed a Rutherford trajectory. 
The TDHF equation is then solved  iteratively in time with a time step $\Delta t= 0.2$~fm$/$c.
The single particle occupation numbers are kept constant in the time evolution according to the frozen occupation approximation (FOA). 
The FOA allows to account for static pairing correlations in the initial state, thus providing reasonable deformation of the collision partners. 
However, it neglects dissipative effects from pairing dynamics as well as effects coming from the difference between gauge angles of collision partners \cite{ebata2014a,hashimoto2016,magierski2017}.
Though it is beyond the scope of this work, it would be interesting to investigate the effect of pairing dynamics on quasi-fission in the future. 


The grid size of the TDHF simulations were set to be 28$\times$28$\times$84~fm$^3$ for central collisions and 84$\times$28$\times$84~fm$^3$ for non-central collisions.
The simulations were stopped when the system re-separated or when the contact time (defined in Sec.~\ref{sec:contact}) exceeded 35 zs.
Both static initial conditions and dynamical evolutions are computed with the Sky3D solver~\cite{maruhn2014,schuetrumpf2018}.
The central (non-central) collisions were run for up to 6 (16) days on 8 CPU cores per system.
Multiple systems were simulated in parallel on the Australian National Computational Infrastructure's (NCI) Gadi supercomputer.

To search for quasi-fission, central collisions between pairs of nuclei at energies ranging from 0.9~$V_B$ to 1.3~$V_B$ were simulated in 20 equal steps in energy, where $V_B$ is the Coulomb barrier of the system.
A finer energy search was run until the fusion threshold could be determined within 0.1~MeV. 
The barriers were determined according to \'{S}wiatecki \textit{et al}. \cite{swiatecki2005}, and found to be $V_B=151.8$~MeV and $215.4$~MeV  for $^{50}$Ca$+^{176}$Yb and $^{96}$Zr$+^{130}$Sn, respectively.
Non-central collisions were simulated at fixed energies of 10\% above their Coulomb barriers (167~MeV and 237~MeV, respectively) to investigate the effect of angular momentum.
The results are summarised in the Appendix.

\section{TDHF Results}\label{sec:results}

\subsection{Quasi-fission properties}

\subsubsection{Contact times\label{sec:contact}}
TDHF simulations of heavy-ion collisions do not enforce quasi-fission to occur, so simulation outcomes that have led to quasi-fission must be searched for.
Characteristics of quasi-fission include full damping of kinetic energy of the collision partners, significant mass transfer, and contact times exceeding few zeptoseconds. 
In the following, contact times are used to define criteria for characterisation of the main reaction mechanisms. 
Here, two nuclei are considered to be in contact when the neck density exceeds 0.08~fm$^{-3}$, i.e., half of the nuclear saturation density.

The timescale for kinetic energy dissipation being of the order of 2 zs \cite{simenel2021}, 
we consider that reactions with contact times $\tau<2$~zs are not fully damped and are thus not associated with quasi-fission. 
Such reactions are generically called hereafter ``quasi-elastic'' scattering (QS), though they also include deep-inelastic scattering. 
Systems with contact times greater than 30~zs were usually associated\footnote{In some cases where the system's elongation was  increasing at $\tau\simeq35$~zs, simulations were run for a longer time.} with ``fusion'', although, strictly speaking, these  include potential slow quasi-fission as well.

\begin{figure}[!htb]
    \centering
    \includegraphics[width=\linewidth]{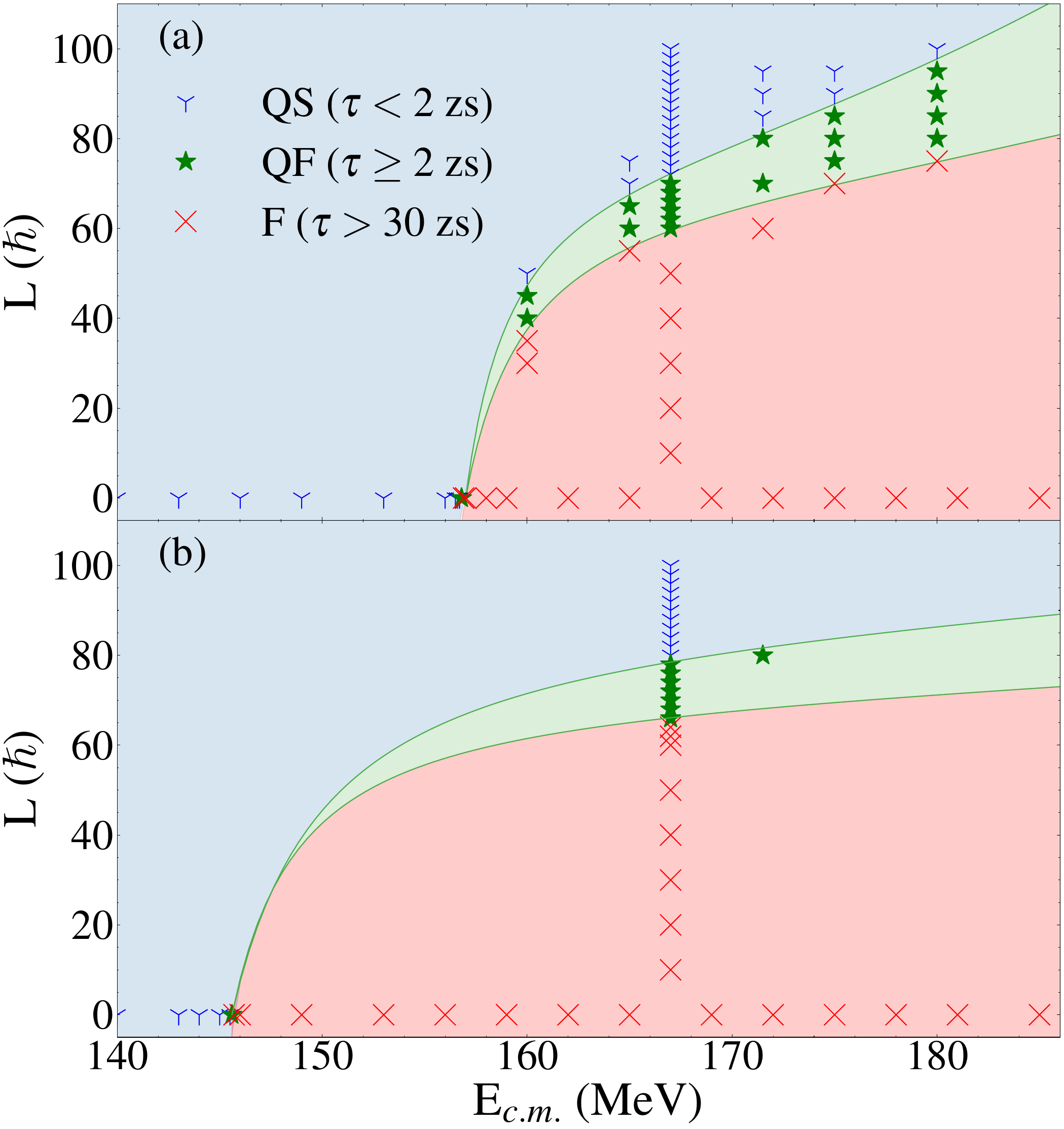}
    \caption{Simulated results for side-oriented (a) and tip-oriented (b) $^{50}$Ca$+^{176}$Yb. Each point corresponds to a TDHF simulation. 
    The blue, green and red shaded regions correspond to regions assigned to ``quasi-elastic scattering'' (QS), ``quasi-fission'' (QF) and ``fusion'' (F), respectively.
 The region boundaries are only approximate guidelines, obtained by fitting a quadratic polynomial in inverse energy.}
    \label{fig:LE}
\end{figure}

$^{50}$Ca$+^{176}$Yb central collisions are characterised by a rapid transition from quasi-elastic scattering to fusion (see $L=0$ entries in Tabs.~\ref{tab:tip_Yb} and~\ref{tab:side_Yb} in the Appendix). 
For each orientation, only one re-separation occurred at $\tau>2$~zs contact time: $\tau=2.75$~zs at $E_{c.m.}=145.6$~MeV (tip) and $\tau=2.09$~zs at $E_{c.m.}=156.8$~MeV (side). 
In each case, less than one nucleon is transferred in average. The relatively short contact times and small mass transfer indicate that these reactions could still be considered as QS.  
With only 0.1~MeV increase in energy, these systems fuse with contact times exceeding 35~zs. 
We conclude that no quasi-fission occurs in TDHF simulations of  $^{50}$Ca$+^{176}$Yb central collisions.
This is in contrast with heavier systems \cite{oberacker2014,mcglynn2023}. 

The simulation results for $^{50}$Ca$+^{176}$Yb collisions with the side orientation at various angular momenta and centre-of-mass energies are presented in Fig. \ref{fig:LE}(a).
Interestingly, quasi-fission is observed at finite orbital angular momentum in the transition between fusion and QS. 
A similar behaviour was also seen with the tip oriented collisions in Fig. \ref{fig:LE}(b).

\begin{figure}[!htb]
    \centering
    \includegraphics[width=\linewidth]{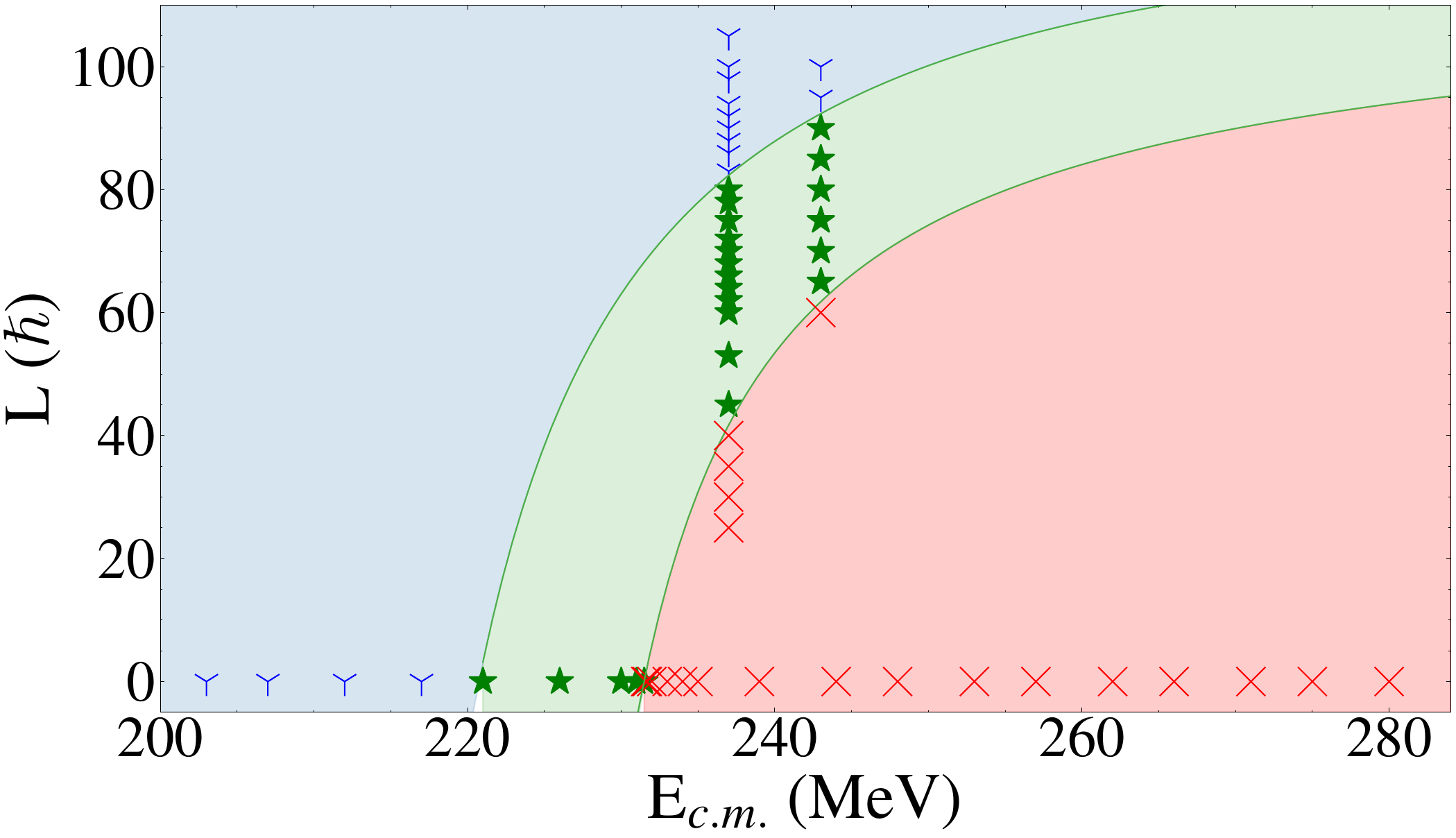}
    \caption{Simulated results for $^{96}$Zr$+^{130}$Sn. Each point corresponds to a TDHF simulation. 
    The blue, green and red shaded regions correspond to regions assigned to ``quasi-elastic scattering'' (QS), ``quasi-fission'' (QF) and ``fusion'' (F), respectively.
    The region boundaries are only approximate guidelines, obtained by fitting a quadratic polynomial in inverse energy.}
    \label{fig:LE2}
\end{figure}

In contrast with $^{50}$Ca$+^{176}$Yb, quasi-fission occurs in $^{96}$Zr$+^{130}$Sn central collisions as indicated by contact times exceeding 2~zs  between $E_{c.m.}\approx221$~MeV and $E_{c.m.}\approx232$~MeV as shown in Fig. \ref{fig:LE2}.
These contact times keep increasing in central collisions up to $\tau\simeq9$~zs  at $E_{c.m.}\approx231$~MeV. 
These energies are above the Coulomb barrier $V_B=215.4$~MeV for this system, indicating a fusion hindrance mechanism. 
However, no re-separation with contact times longer than $10$~zs was observed in this system, even in non-central collisions.

\subsubsection{Mass transfer}
\begin{figure}[!htb]
    \centering
    \includegraphics[width=\linewidth]{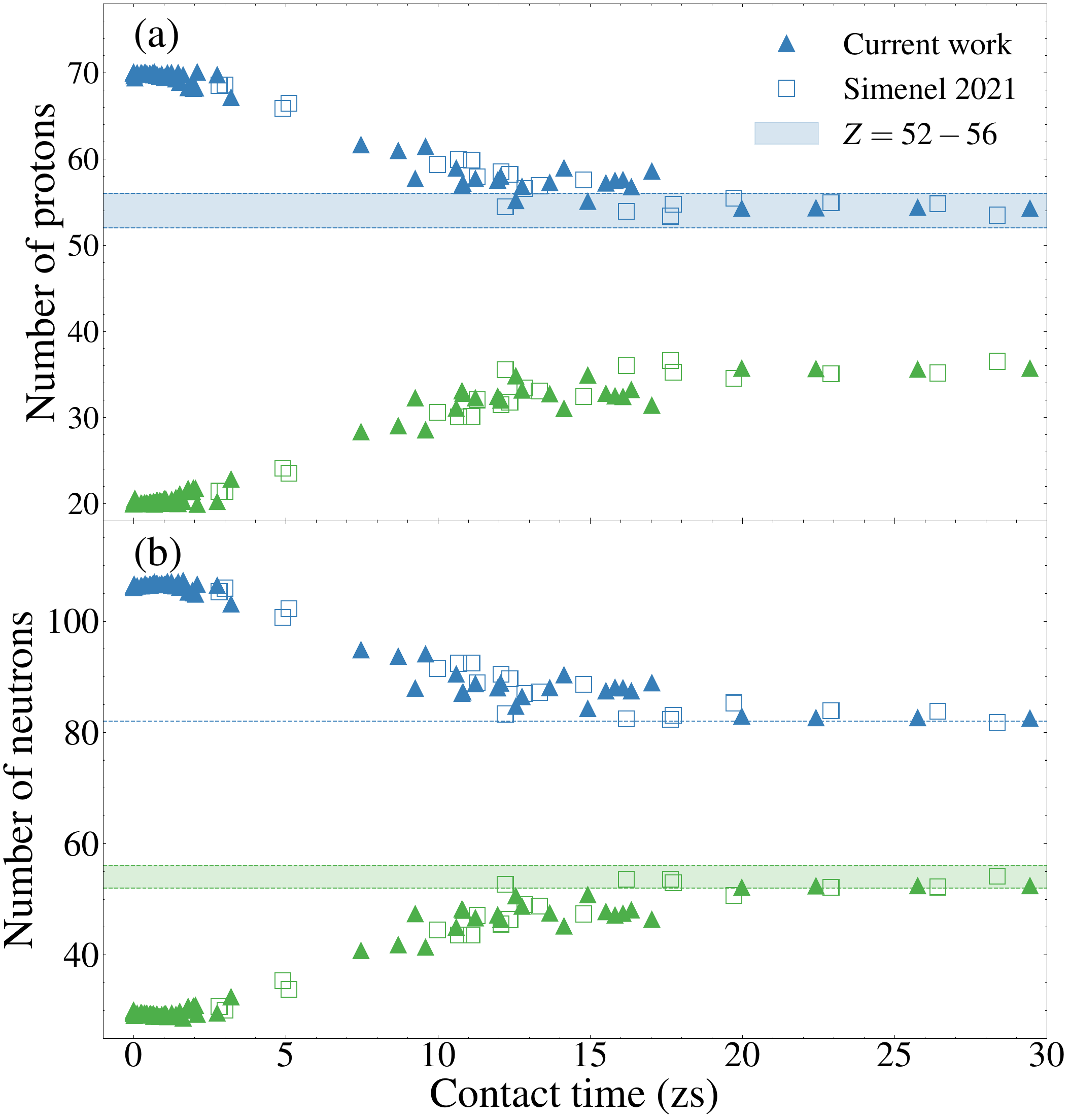}
    \caption{Numbers of (a) protons and (b) neutrons in the outgoing fragments of $^{176}$Yb$+^{50}$Ca collisions against contact time, for both side and tip orientations.
    The blue and the green symbols correspond to the heavy and light fragments, respectively. 
    Present results are shown as triangles while those of Ref.~\cite{simenel2021} (Simenel 2021) are shown as squares.
    The shaded regions at (a) $Z=52-56$ and (b) $N=52-56$ indicate expected octupole deformed shell effects.
    The line in (b) indicates the spherical shell effect at $N=82$.}
    \label{fig:side_Yb176Ca50_Z}
\end{figure}

Figure~\ref{fig:side_Yb176Ca50_Z} shows the evolution of the number of protons in the fragments with contact time for both side and tip orientations of $^{50}$Ca$+^{176}$Yb collisions.
The results agree well with those of Ref.~\cite{simenel2021} that were obtained with the TDHF3D code~\cite{kim1997}.
An increase in the mass transfer is observed with increase in the contact time. 
However, this mass equilibration process stops for contact times greater than $\approx13$~zs, where the numbers of protons (neutrons) converge to $Z_L\approx36$ ({$N_L\approx 52$}) and $Z_H\approx54$ ({$N_H\approx 82$}) in the light and heavy fragments, respectively, as in $^{226}$Th asymmetric fission. 
This could be attributed to several shell effects in the fragments, including octupole deformed shell effects at $Z=52,56$~\cite{simenel2021} that were proposed to explain the final asymmetry in fission of actinides~\cite{scamps2018}. 

\begin{figure}[!htb]
\centering
\includegraphics[width=\linewidth]{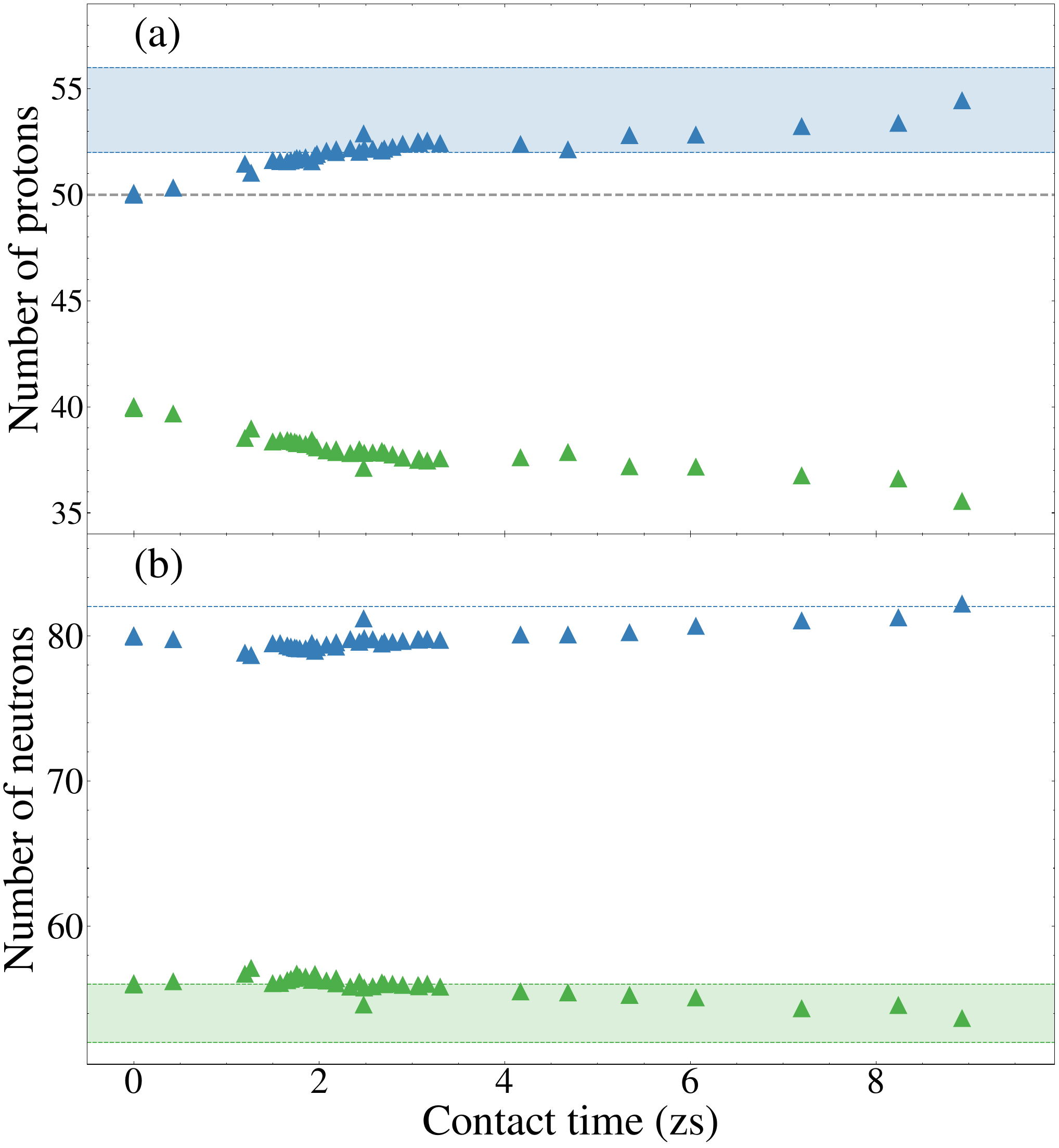}
\caption{Numbers of (a) protons and (b) neutrons in the outgoing fragments of $^{96}$Zr$+^{130}$Sn collisions against contact time.
    The blue and the green symbols correspond to the heavy and light fragments, respectively. 
The shaded regions at $Z,N=52-56$ indicate expected octupole deformed shell effects. The lines at $Z=50$ and $N=82$ indicate nuclear magic numbers.}
\label{fig:Zr96Sn130_Z}
\end{figure}

The numbers of protons and neutrons in the outgoing fragments produced in  $^{96}$Zr$+^{130}$Sn collisions are plotted in Figs.~\ref{fig:Zr96Sn130_Z}(a) and~\ref{fig:Zr96Sn130_Z}(b), respectively, as a function of contact time. 
While the numbers of neutrons in the fragments remain close to those of the collision partners, some protons are transferred from $^{96}$Zr to $^{130}$Sn, moving the latter away from magicity associated with spherical shell gap at $Z=50$. 
This transfer occurs within $\approx2$~zs, a timescale characteristic to neutron-to-proton equilibration \cite{jedele2017,umar2017,simenel2020}.
Indeed, the collision partners have $N/Z=1.4$ ($^{96}$Zr) and $1.6$ ($^{130}$Sn) while the fragments exiting the reaction after $\tau\approx2$~zs contact time have $N/Z\approx1.5$. 
The net effect is a larger mass asymmetry in the exit channel than in the entrance one. This asymmetry is slightly increased for longer contact times. 
This unusual drift away from mass symmetry is referred to as ``inverse quasi-fission''.
Inverse quasi-fission was also observed in other systems \cite{zagrebaev2006,kedziora2010,kozulin2017} and attributed to orientation or shell effects.
As a result, the outgoing fragments formed in  $^{96}$Zr$+^{130}$Sn are similar to those produced both in $^{176}$Yb$+^{50}$Ca collisions and in $^{226}$Th asymmetric fission, indicating, once again, a possible influence of octupole deformed shell effects in the fragments fixing their final asymmetry. 

\subsubsection{Total kinetic energy}

\begin{figure}[!htb]
    \centering
    \includegraphics[width=\linewidth]{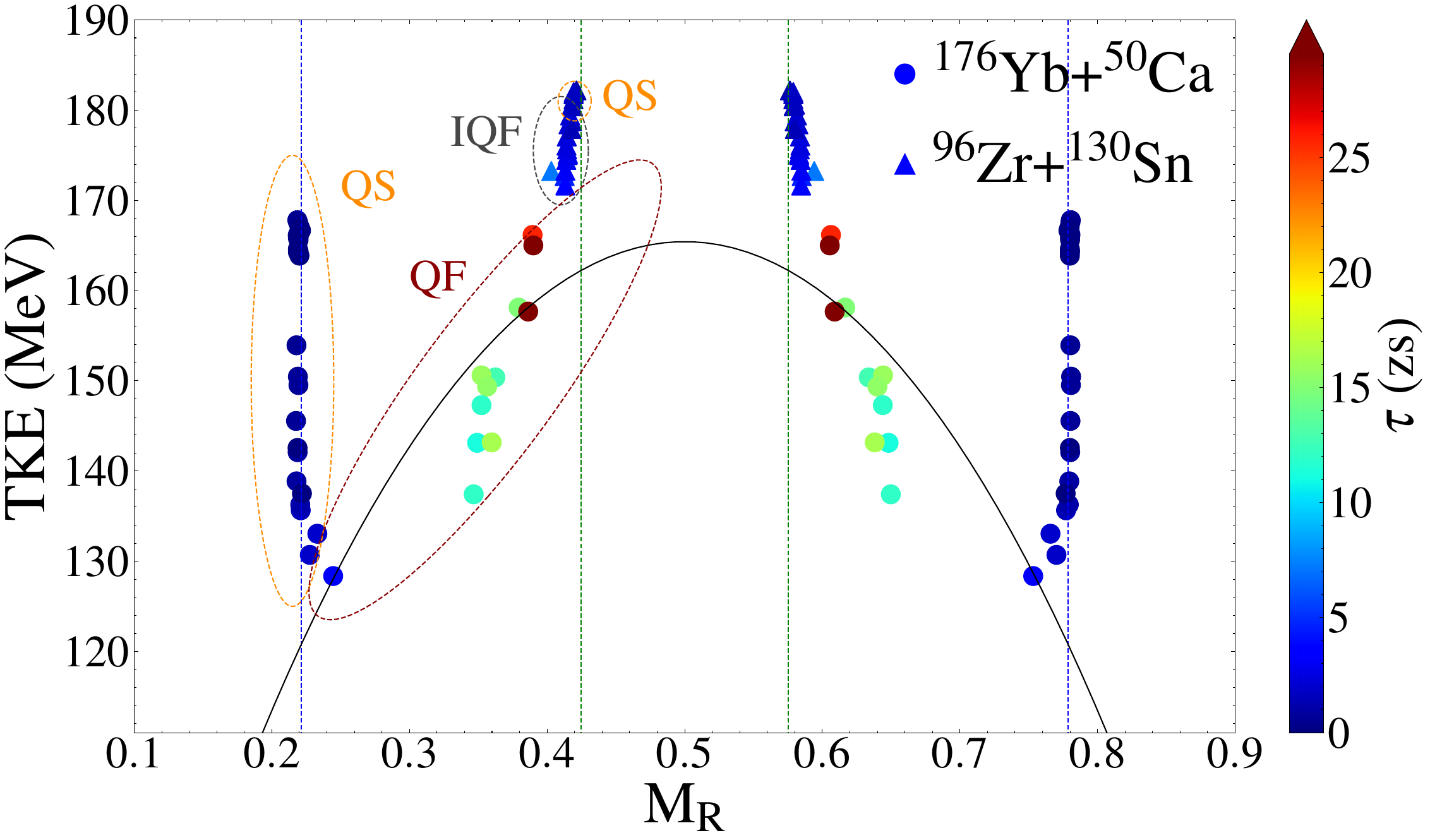}
    \caption{Total kinetic energy of fragments from collisions of $^{176}$Yb$+^{50}$Ca at $E_{c.m.}=167$~MeV and $^{96}$Zr$+^{130}$Sn at $E_{c.m.}=237$~MeV as a function of mass ratio $M_R$. 
    Quasi-elastic scattering (QS), quasi-fission (QF), and inverse quasi-fission (IQF) events are indicated for the light fragments.
    The blue and green vertical dashed lines show the initial mass ratio for the reacting systems of $^{176}$Yb$+^{50}$Ca and $^{96}$Zr$+^{130}$Sn, respectively. 
    The total kinetic energy from Viola systematics \cite{viola1985,hinde1987} is shown as the black solid line.
    The color scale gives the contact time $\tau$ of each reaction. }
    \label{fig:TKE_Viola_Yb}
\end{figure}

Figure~\ref{fig:TKE_Viola_Yb} shows the TKE of the outgoing fragments as a function of the mass ratio $M_R=A_{frag.}/A_{tot.}$ where $A_{frag.}$ is the fragment mass number and $A_{tot.}=226$ is the total one. 
The quasi-fission events have TKE close to Viola systematics~\cite{viola1985,hinde1987}, confirming that these are fully damped events. 
However, the inverse quasi-fission events occurring in the $^{96}$Zr$+^{130}$Sn reaction have TKE  higher than Viola systematics. 
This could be attributed to the proximity of $Z=50$ spherical shell gap that is expected to produce more compact fragments. 
In particular, the $Z=52$ octupole deformed shell gap is less deformed (thus leading to more compact configurations at scission and therefore to higher TKE) than the $Z=56$ one. 
Interestingly, the most symmetric events produced  in $^{176}$Yb$+^{50}$Ca collisions seem to converge towards the same region of the TKE$-M_R$ plot as the most asymmetric outgoing fragments formed in $^{96}$Zr$+^{130}$Sn  inverse quasi-fission, indicating that the fragments are likely to be produced with similar shapes. 

\subsubsection{Total excitation energy}

\begin{figure}[!htb]
    \centering
    \includegraphics[width=\linewidth]{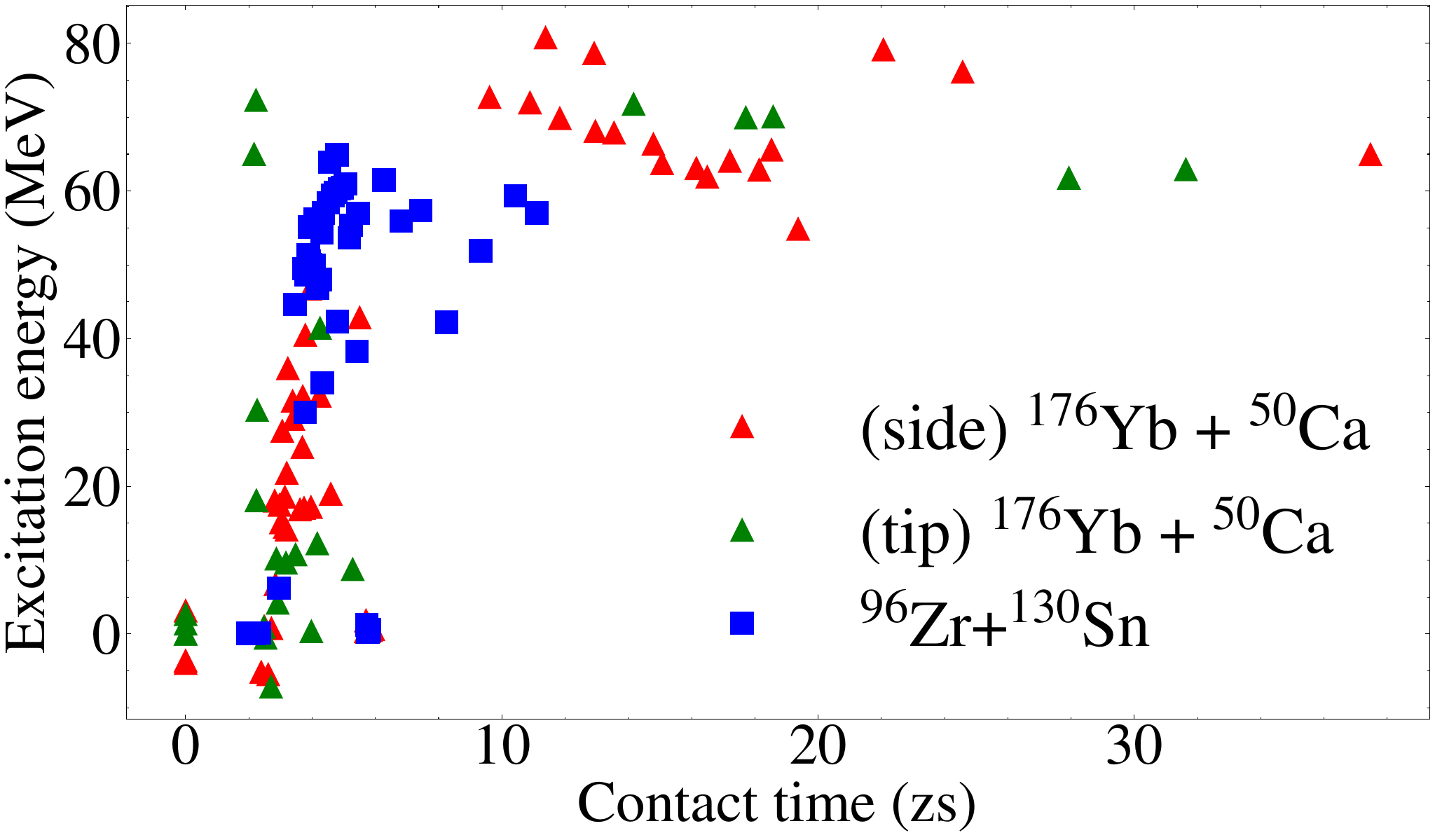}
    \caption{Total excitation energy (TXE) of the simulated systems as function of contact times. The red and green symbols correspond to collisions between $^{50}$Ca and $^{176}$Yb in side and tip orientations, respectively. The blue symbols correspond to $^{96}$Zr+$^{130}$Sn collisions.}
    \label{fig:Ex}
\end{figure}

The total excitation energy (TXE) in the exit channel can be evaluated from $TXE = Q + E_{c.m.}-TKE$, where $Q$ is the $Q-$value for the specific reaction channel.
$Q-$values were derived using Ref.~\cite{wang2021}.
Figure~\ref{fig:Ex} shows a rapid increase of TXE with contact time, followed by a plateau at $TXE\simeq70\pm10$~MeV at $\tau\gtrsim5$~zs. 
The results are obtained for various energies and impact parameters, indicating little dependence of quasi-fission outcome with energy and angular momentum. 

This TXE is shared by the outgoing fragments (See Ref.~\cite{umar2016} for an evaluation of this repartition in other systems from the density-constrained TDHF method).
The resulting excitation energy at scission could, in principle, be large enough to wash out shell effects. 
It should be noted, however, that the systems are not thermalised at scission and a significant fraction of the excitation energy is expected to be stored into deformation energy and collective modes. Evaluating the repartition between different types of excitation energy is an interesting prospect for future studies. 

\section{Quasi-fission trajectories}\label{sec:compare}

Let us know compare TDHF trajectories in the $Q_{20}-Q_{30}$ plane with the PES, as in Ref.~\cite{mcglynn2023}. 
This is done in Figs.~\ref{fig:side_YbCa_all} and~\ref{fig:YbCa_dep} for $^{50}$Ca$+^{176}$Yb central and non-central collisions, respectively, and in Fig.~\ref{fig:ZrSn_Edep} for $^{96}$Zr$+^{130}$Sn central collisions. 
Each trajectory can be divided into two parts: the entrance channel trajectory where the two colliding nuclei are yet to touch (dashed line) and the following trajectory where the nuclei are in contact (solid line).
The entry point where the two nuclei first make contact is represented by a star.

\subsection{$^{50}$Ca$+^{176}$Yb}

\begin{figure}[!htb]
    \centering
    \includegraphics[width=\linewidth]{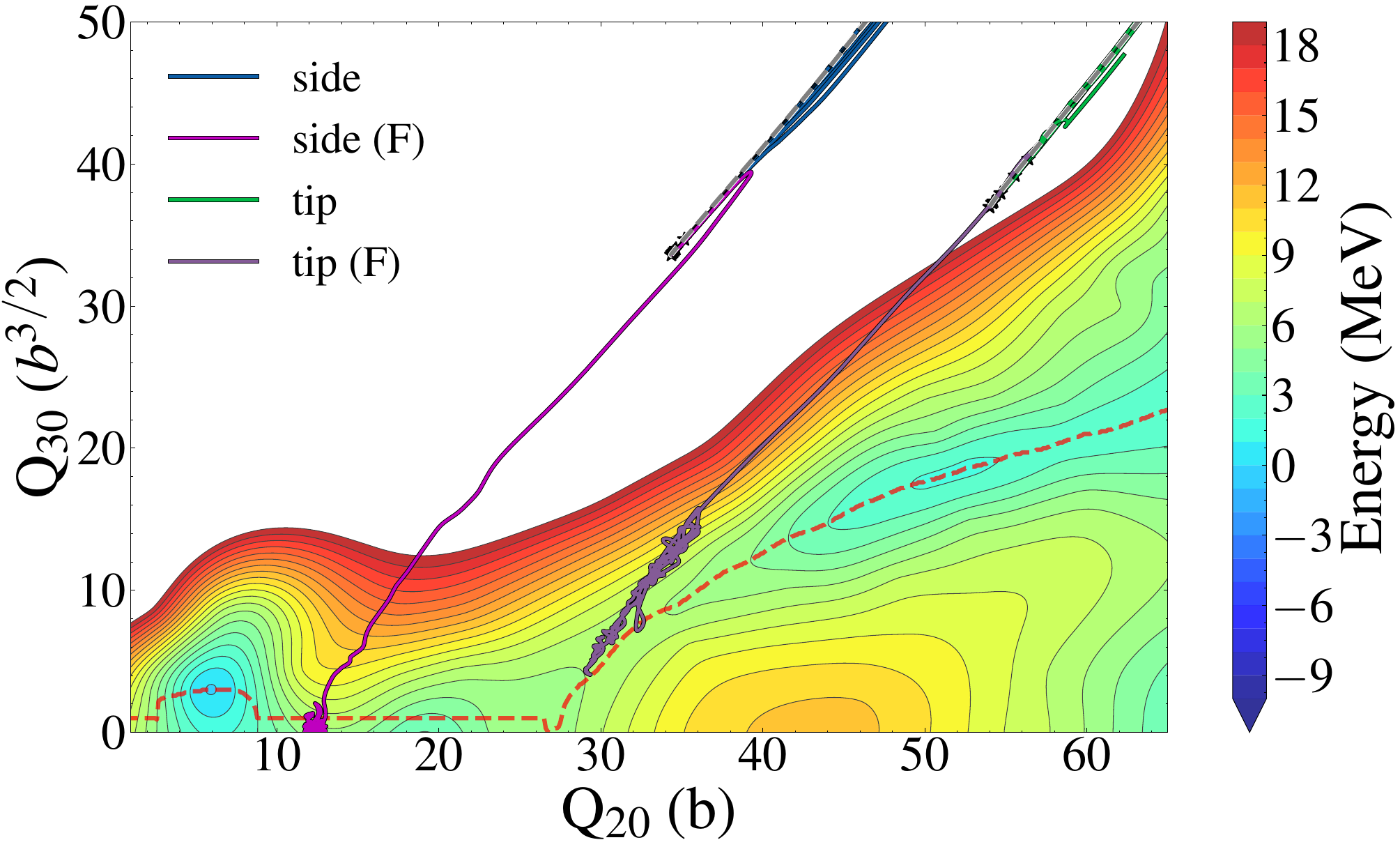}
    \caption{ TDHF trajectories of $^{50}$Ca$+^{176}$Yb central collisions at $E_{c.m.}=140-145.7$~MeV ($E_{c.m.}=153-156.9$~MeV) with tip (side) orientation are drawn in the $Q_{20}-Q_{30}$ plane on top of the PES.  Trajectories leading to ``fusion''  are denoted by (F). The entrance channel trajectories  are shown by the blacked dashed lines and the entry (contact) point are represented by stars. The colored solid lines represent trajectories after contact. The orange dashed line shows the minimum energy fission path.}
    \label{fig:side_YbCa_all}
\end{figure}

As discussed earlier, $^{50}$Ca$+^{176}$Yb central collisions do not lead to quasi-fission. 
The $Q_{20}-Q_{30}$ TDHF trajectories that lead to a re-separation in Fig.~\ref{fig:side_YbCa_all} then follow closely the entrance channel trajectory, which is compatible with quasi-elastic scattering.
Above the fusion threshold, the system then drifts toward the formation of a more compact system. 
In particular, the side orientation, which is more compact at the entry point, evolves toward a shape with $Q_{30}\approx0$ and an elongation comparable to that of the first fission barrier. 
Note that central collisions with side orientations lead to non-axial shapes. 
Thus using features of the axial PES to interpret TDHF trajectories for side orientations should be done with care.
Central collisions with the tip orientation, however, should preserve their entrance channel axial symmetry. 
In this case, the PES is relevant to interpret the TDHF trajectories, assuming that excitation energy does not affect the PES topography. 
We see in Fig.~\ref{fig:side_YbCa_all} that the tip orientation leading to fusion evolves toward a shape with elongation and asymmetry compatible with the second barrier (or saddle point). 
It is interesting to see that both ``fusion'' trajectories in Fig.~\ref{fig:side_YbCa_all} lead to configurations close to the minimum energy  fission path.  
The tip orientation, in particular, might as well follow this path to scission, leading to a so-called ``slow'' quasi-fission. 

\begin{figure}[!htb]
    \centering
    \includegraphics[width=\linewidth]{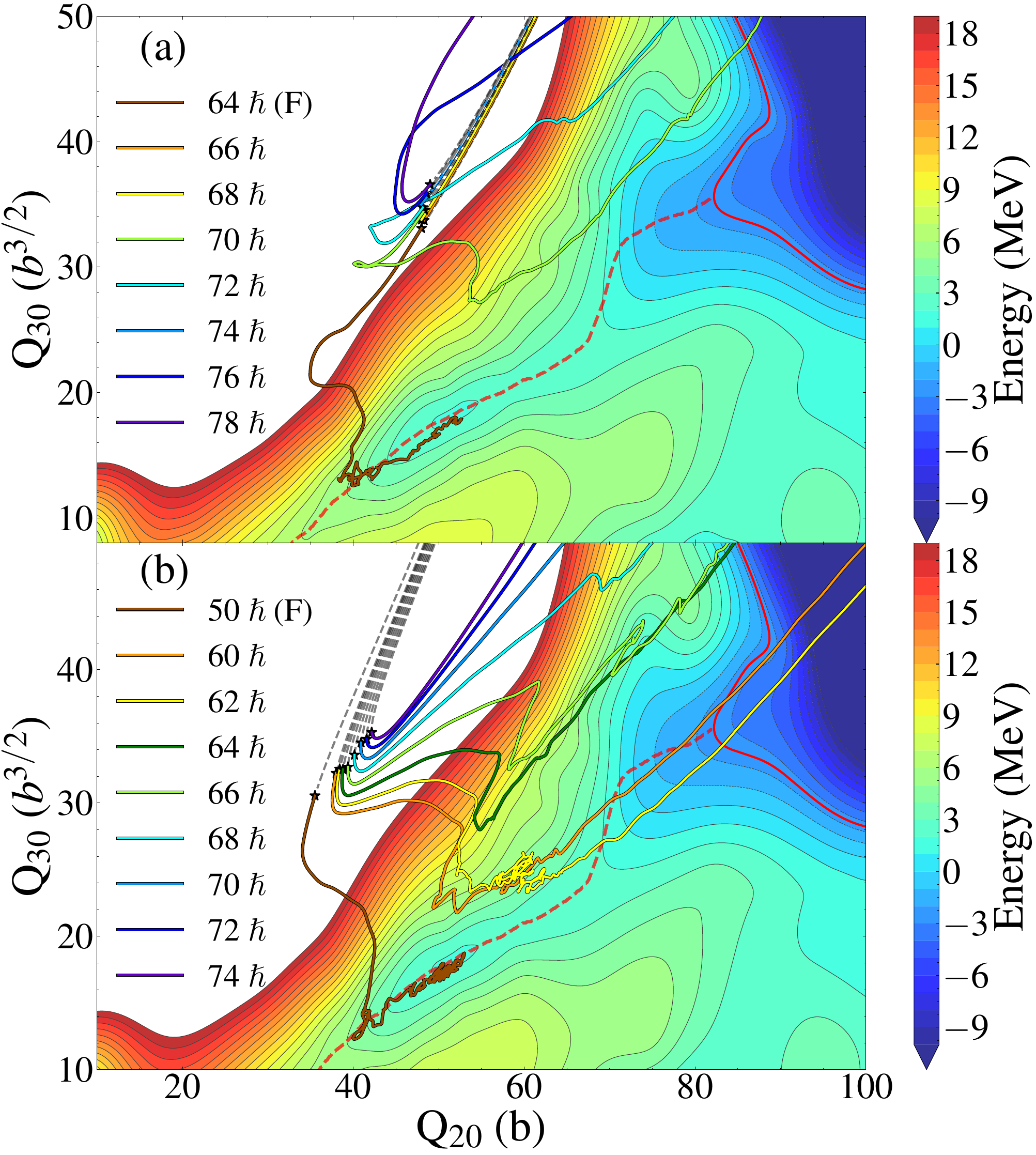}
    \caption{Same as Fig.~\ref{fig:side_YbCa_all} for quasi-fission trajectories of $^{50}$Ca$+^{176}$Yb at $E_{c.m.}=167$~MeV and various angular momenta for tip (a) and side (b) orientations.}
    \label{fig:YbCa_dep}
\end{figure}

Figure~\ref{fig:YbCa_dep}(a) and \ref{fig:YbCa_dep}(b) show the TDHF trajectories for $^{50}$Ca$+^{176}$Yb at $E_{c.m.}=167$~MeV at finite $L$ with the tip and side orientations, respectively. 
The underlying PES being computed at $L=0$ and assuming axial symmetry, it only provides a qualitative tool for comparing with TDHF trajectories which, for non-central collisions, all break axial symmetry.

Nevertheless, it is interesting to see that the less peripheral collisions that still lead to quasi-fission do it by following the  asymmetric fission valley. 
Moreover, the trajectories that are associated with ``fusion'', seem in fact to get trapped into a local minimum along the minimum energy fission path. 
This could be an indication that some features of the PES remain relevant at finite $L$ and for non-axial shapes, or that the system evolves towards approximately axial shapes during the reaction.

\begin{figure}[!htb]
    \centering
    \includegraphics[width=\linewidth]{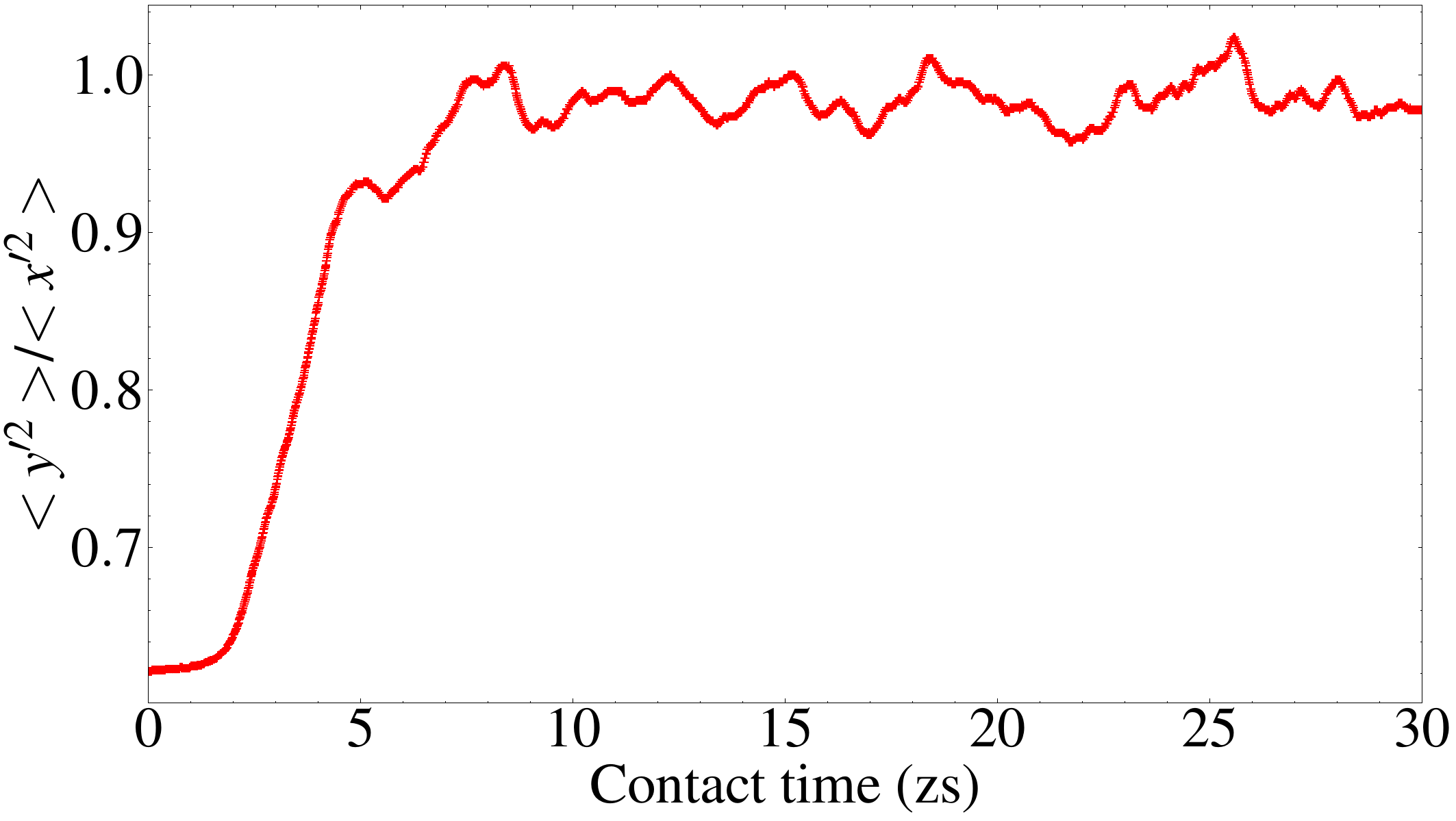}
    \caption{Time evolution of $\langle y'^2\rangle/\langle x'^2\rangle$ (see text for definition) for $^{50}$Ca+$^{176}$Yb (side orientation) at $E_{c.m.}=167$~MeV and $L = 50\hbar$.}
    \label{fig:xy}
\end{figure}

To get a deeper insight into the shape evolution of the system, the principal axis $z'$ was determined from diagonalisation of the cartesian quadrupole tensor. 
The $^{176}$Yb deformation axis being initially in the collision plane $(x,z)$, the $z'$ axis also remains in the collision plane (so does, by definition, the $x'$ axis). 
Thus, for an axially symmetric system, one would expect $\langle x'^2\rangle=\langle y'^2\rangle$.
Figure~\ref{fig:xy} shows the evolution of $\langle y'^2\rangle/\langle x'^2\rangle$ for the $^{50}$Ca+$^{176}$Yb reaction with the side orientation at $E_{c.m.}=167$~MeV and $L = 50\hbar$ that seems to get ``trapped'' into a  pocket of the PES in Fig.~\ref{fig:YbCa_dep}(b) (solid brown line). 
We see that this ratio is initially smaller than 1, indicating a non-axial shape. However, it rapidly increases and become approximatively constant with $\langle y'^2\rangle/\langle x'^2\rangle\simeq1$, compatible with a quasi-axial shape. 
In this case, the comparison between the trajectory and the PES topography is meaningful despite the initial non-axiality. 
Note that not all systems are guaranteed to evolve towards an axial shape (see, e.g., Ref.~\cite{mcglynn2023}). 
Thus the shape of the system should be studied (e.g., through the ratio $\langle y'^2\rangle/\langle x'^2\rangle$ as in the example of Fig.~\ref{fig:xy}) whenever a comparison with the PES is relevant.

\subsection{$^{96}$Zr$+^{130}$Sn}

\begin{figure}[!htb]
    \centering
    \includegraphics[width=\linewidth]{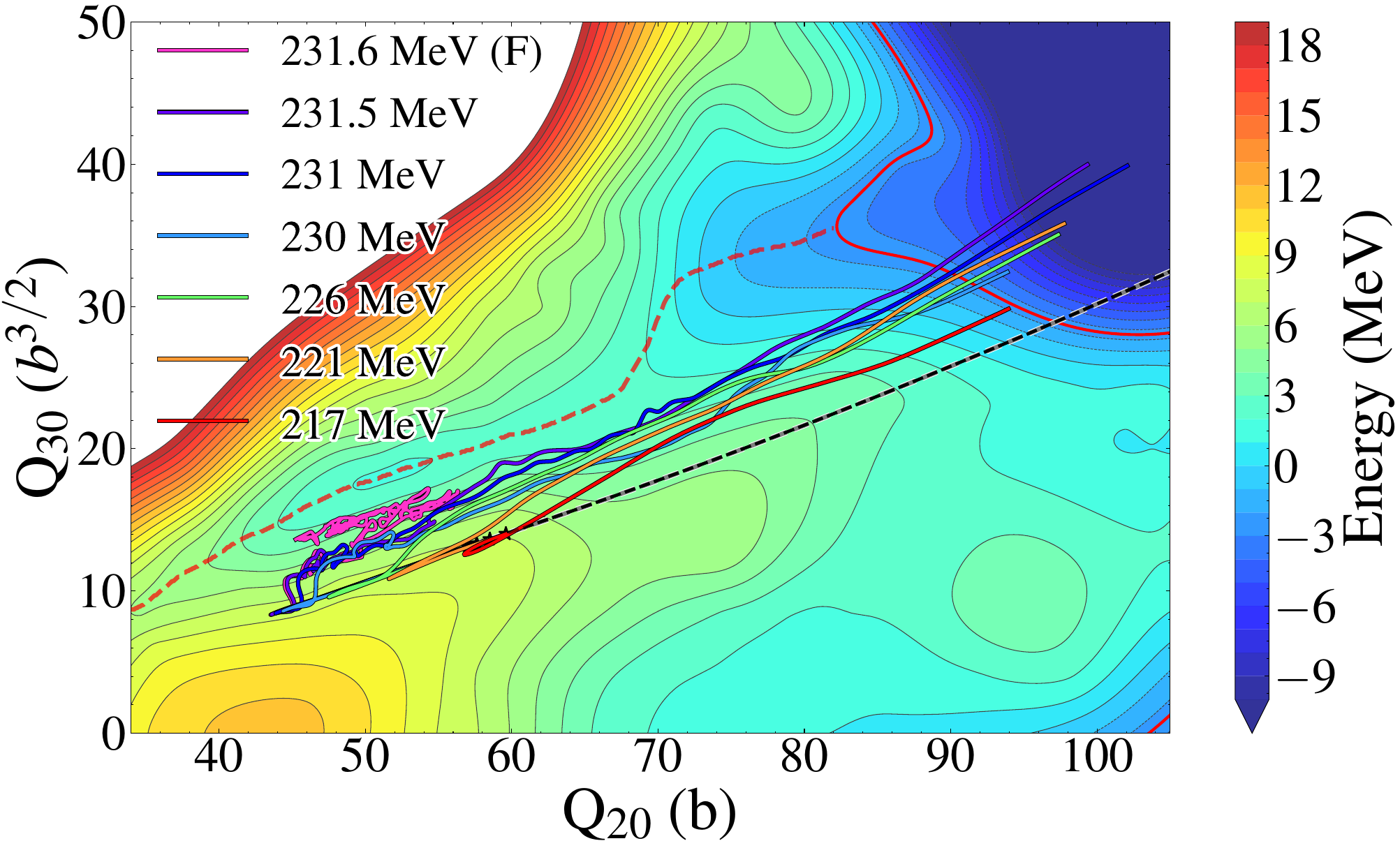}
    \caption{Same as Fig.~\ref{fig:side_YbCa_all} for $^{96}$Zr$+^{130}$Sn central collisions at $E_{c.m.}=217-235$~MeV.}
    \label{fig:ZrSn_Edep}
\end{figure}

Figure~\ref{fig:ZrSn_Edep} shows the TDHF trajectories in $^{96}$Zr$+^{130}$Sn central collisions\footnote{Non-central collisions, not shown in Fig.~\ref{fig:ZrSn_Edep}, exhibit a similar behaviour. See also Tab.~\ref{tab:Zr} in the Appendix.}. 
After contact, the trajectories follow more asymmetric exit channels leading to inverse quasi-fission.
Axial symmetry should be preserved in these reactions and thus a comparison with the underlying PES is meaningful. 
Interestingly, the trajectories all seem to ``hit'' the second barrier, preventing all but the highest energy to fuse. 
As discussed earlier, proton-to-neutron equilibration may explain the initial increase in asymmetry. 
However, the presence of the asymmetric fission valley seems to further drive the systems towards inverse quasi-fission. 

\section{Discussion and conclusions}\label{sec:conc}
The influence of shell effects in quasi-fission was studied with TDHF simulations of $^{50}$Ca$+^{176}$Yb and $^{96}$Zr$+^{130}$Sn reactions. The trajectories in the $Q_{20}-Q_{30}$ plane were  compared to the PES of the $^{226}$Th compound nucleus.
Mass equilibration in the $^{50}$Ca$+^{176}$Yb reaction stops when the system reaches  the asymmetric fission valley of $^{226}$Th, leading to heavy fragments with $Z\approx54$ protons.
This is an indication that quasi-fission can be affected by the same shell effects as in fission. In this case, the octupole deformed shell effects at $Z=52$ and $56$ are invoked as a factor impacting the final asymmetry of the system~\cite{scamps2018}. 

The exit trajectories of $^{96}$Zr$+^{130}$Sn are found to be more mass-asymmetric than the entrance channel ones, contrary to the expectation based on a simple liquid drop model of quasi-fission which usually drives the system towards fragment mass symmetry.
The rapid neutron-to-proton equilibration may explain an initial drive towards asymmetry. 
Although the TDHF trajectories do not follow closely the bottom of the asymmetric fission fission valley, the latter seem to further drive the systems towards inverse quasi-fission. 
No trajectories were found to enter the symmetric fission valley in $^{226}$Th PES for this reaction. 
As such, the observation of more mass-asymmetric $^{96}$Zr$+^{130}$Sn trajectories, potentially affected by the asymmetric fission valley, further suggests that quasi-fission is subject to shell effects.

In many situations, the comparison between TDHF trajectories and PES remains qualitative as the PES is computed assuming zero excitation energy, zero angular momentum, and  axial symmetry. 
Each of these assumptions is expected to break down in heavy-ion collisions. 
It would therefore be interesting to compute PES without these assumptions in order to investigate their effects on the PES topography, and in particular, on the fission valleys induced by shell effects.

\begin{acknowledgments}

This work has been supported by the Australian Research Council Discovery Project (project number DP190100256).
Computational resources were provided by the Australian Government through the National Computational Infrastructure (NCI) under the ANU Merit Allocation Scheme, the Adapter scheme, and the National Committee Merit Allocation Scheme. H. Lee acknowledges the support of the Australian National University through the Dunbar Physics Honours scholarship.
P. McGlynn acknowledges the support of the Australian National University through the Deakin PhD scholarship and the Dean’s Merit Scholarship in science.

\end{acknowledgments}

\appendix
\section*{APPENDIX: TDHF Simulation results}

\FloatBarrier

\begin{table}[!ht]
    \caption{Exit channel properties of $^{96}$Zr$+^{130}$Sn. Energies ($E_{c.m.}$ and TKE) are in MeV, angular momenta ($L$) in units of $\hbar$, and contact times ($\tau$) in zeptoseconds (zs). Subscripts $H$ and $L$ refer to heavy and light fragments, respectively.}\label{tab:Zr}
\begin{ruledtabular}
    \begin{tabular}{cccccccc}
        $E_{c.m.}$ & $L$ & $\tau$ & $A_H$ & $A_L$ & $Z_H$ & $Z_L$ & TKE \\ \hline 
        203 & 0 & 0.00 & 129.99 & 95.98 & 50.08 & 39.92 & 201.76 \\ 
        207 &  & 0.43 & 130.07 & 95.87 & 50.33 & 39.67 & 200.81 \\ 
        212 &  & 1.27 & 129.68 & 96.07 & 51.02 & 38.96 & 177.02 \\ 
        217 &  & 1.92 & 131.04 & 94.73 & 51.56 & 38.43 & 177.74 \\ 
        221 &  & 2.48 & 134.06 & 91.71 & 52.88 & 37.11 & 171.91 \\ 
        226 &  & 3.16 & 132.29 & 93.47 & 52.54 & 37.45 & 172.39 \\ 
        230 &  & 6.06 & 133.49 & 92.25 & 52.82 & 37.17 & 174.75 \\ 
        231 &  & 8.24 & 134.62 & 91.17 & 53.38 & 36.61 & 167.63 \\ 
        231.5 &  & 8.93 & 136.64 & 89.21 & 54.44 & 35.55 & 167.82 \\ 
        231.6 &  & $>35$ &  & &  &  &  \\ 
        \hline
        237 & 40 & $>30$ & & & & & \\
         & 45 & 7.20 & 134.26 & 91.09 & 53.22 & 36.75 & 173.24 \\ 
         & 53 & 4.68 & 132.20 & 93.28 & 52.13 & 37.85 & 172.69 \\ 
         & 60 & 3.30 & 132.11 & 93.38 & 52.42 & 37.56 & 171.64 \\ 
         & 62 & 3.08 & 132.14 & 93.40 & 52.43 & 37.55 & 173.27 \\ 
         & 64 & 2.90 & 132.02 & 93.55 & 52.39 & 37.60 & 174.48 \\ 
         & 66 & 2.79 & 131.79 & 93.76 & 52.24 & 37.74 & 175.01 \\ 
         & 68 & 2.70 & 131.75 & 93.81 & 52.16 & 37.82 & 175.16 \\ 
         & 70 & 2.58 & 131.88 & 93.68 & 52.15 & 37.83 & 175.64 \\ 
         & 72 & 2.48 & 131.98 & 93.58 & 52.17 & 37.81 & 176.07 \\ 
         & 75 & 2.33 & 131.93 & 93.63 & 52.18 & 37.80 & 177.08 \\ 
         & 78 & 2.18 & 131.66 & 93.91 & 52.13 & 37.85 & 178.42 \\ 
         & 80 & 2.08 & 131.42 & 94.17 & 52.06 & 37.93 & 179.41 \\ 
         & 83 & 1.97 & 131.11 & 94.49 & 51.91 & 38.07 & 180.60 \\ 
         & 86 & 1.85 & 130.85 & 94.74 & 51.75 & 38.23 & 181.77 \\ 
\hline
	243 & 60 & $>30$ & & & & & \\
        & 65 & 5.34 & 133.02 & 92.44 & 52.80 & 37.18 & 172.63 \\ 
         & 70 & 4.17 & 132.44 & 93.11 & 52.38 & 37.60 & 173.15 \\ 
         & 75 & 3.06 & 132.26 & 93.42 & 52.51 & 37.48 & 174.03 \\ 
         & 80 & 2.67 & 131.52 & 94.02 & 52.07 & 37.91 & 176.56 \\ 
         & 85 & 2.43 & 131.58 & 94.14 & 52.01 & 37.98 & 177.54 \\ 
         & 90 & 2.18 & 131.23 & 94.38 & 52.00 & 37.99 & 179.04 \\ 
         & 95 & 1.95 & 130.78 & 94.85 & 51.83 & 38.16 & 181.51 \\ 
         & 100 & 1.76 & 130.85 & 94.99 & 51.72 & 38.28 & 182.62 \\
    \end{tabular}
\end{ruledtabular}
\end{table}

\begin{table}[!ht]
    \caption{Same as Tab.~\ref{tab:Zr} for $^{50}$Ca$+^{176}$Yb(tip)}\label{tab:tip_Yb}
\begin{ruledtabular}
    \begin{tabular}{cccccccc}
        $E_{c.m.}$ & $L$ & $\tau$ & $A_H$ & $A_L$ & $Z_H$ & $Z_L$ & TKE \\ \hline 
        140 & 0 & 0.00 & 176.40 & 49.58 & 70.00 & 20.00 & 138.94 \\ 
        143 &  & 0.38 & 176.66 & 49.29 & 70.02 & 19.98 & 138.01 \\ 
        144 &  & 0.64 & 176.69 & 49.21 & 69.99 & 20.00 & 133.58 \\ 
        145 &  & 0.93 & 176.52 & 49.39 & 69.80 & 20.20 & 133.46 \\ 
        145.5 &  & 1.63 & 177.01 & 48.90 & 69.74 & 20.26 & 132.46 \\ 
        145.6 &  & 2.75 & 176.16 & 49.77 & 69.75 & 20.25 & 136.85 \\ 
        145.7 &  &  $>35$ &  & &  &  &  \\ 
\hline
        167 & 64 & $>35$ &  & &  &  &  \\ 
         & 66 & 25.76 & 137.01 & 88.01 & 54.38 & 35.60 & 166.16 \\ 
         & 68 & 29.45 & 136.81 & 88.13 & 54.26 & 35.71 & 165.00 \\ 
         & 70 & 15.52 & 144.64 & 80.54 & 57.20 & 32.79 & 149.36 \\ 
         & 72 & 11.96 & 145.52 & 79.62 & 57.53 & 32.45 & 147.30 \\ 
         & 74 & 12.06 & 146.84 & 78.34 & 57.98 & 32.01 & 137.42 \\ 
         & 76 & 16.35 & 144.19 & 81.30 & 56.76 & 33.24 & 143.18 \\ 
         & 78 & 2.03 & 173.05 & 52.66 & 68.22 & 21.77 & 133.05 \\ 
         & 80 & 0.04 & 175.54 & 50.13 & 69.39 & 20.59 & 137.51 \\ 
    \end{tabular}
\end{ruledtabular}
\end{table}

\begin{table}[!ht]
    \caption{Same as Tab.~\ref{tab:Zr} for $^{50}$Ca$+^{176}$Yb(side).}\label{tab:side_Yb}
\begin{ruledtabular}
    \begin{tabular}{cccccccc}
        $E_{c.m.}$ & $L$ & $\tau$ & $A_H$ & $A_L$ & $Z_H$ & $Z_L$ & TKE \\ \hline 
        153 & 0 & 0.00 & 176.20 & 49.78 & 70.01 & 19.99 & 151.19 \\ 
        156 &  & 0.68 & 177.02 & 48.88 & 70.05 & 19.95 & 141.09 \\ 
        156.5 &  & 1.11 & 177.02 & 48.89 & 69.96 & 20.04 & 138.87 \\ 
        156.6 &  & 1.25 & 176.99 & 48.92 & 69.98 & 20.02 & 138.75 \\ 
        156.7 &  & 1.47 & 177.00 & 48.90 & 69.99 & 20.01 & 138.73 \\ 
        156.8 &  & 2.09 & 176.67 & 49.23 & 70.07 & 19.93 & 137.05 \\ 
        156.9 &  & $>35$ &  & &  &  &  \\ 
\hline
        160 & 35 &  $>35$ &  & &  &  &  \\ 
         & 40 & 16.08 & 145.55 & 79.86 & 57.57 & 32.42 & 146.50 \\ 
         & 45 & 13.67 & 145.26 & 80.23 & 57.25 & 32.75 & 140.83 \\ 
         & 50 & 1.25 & 176.00 & 49.94 & 69.52 & 20.47 & 134.72 \\ 
\hline
        165 & 55 & $>35$ &  & &  &  &  \\  
         & 60 & 17.03 & 147.47 & 77.78 & 58.57 & 31.42 & 148.15 \\ 
         & 65 & 14.14 & 149.28 & 76.22 & 58.95 & 31.05 & 139.28 \\ 
         & 70 & 1.03 & 176.11 & 49.84 & 69.52 & 20.48 & 135.88 \\ 
         & 75 & 0.77 & 176.40 & 49.49 & 69.75 & 20.25 & 144.16 \\   
\hline
        167 & 50 & $>35$ &  & &  &  &  \\ 
         & 60 & 14.92 & 139.35 & 85.71 & 55.07 & 34.91 & 158.12 \\ 
         & 62 & 35.17 & 137.61 & 87.28 & 54.63 & 35.35 & 157.67 \\ 
         & 64 & 12.76 & 143.22 & 81.90 & 56.81 & 33.18 & 150.38 \\ 
         & 66 & 15.82 & 145.56 & 79.62 & 57.49 & 32.50 & 150.60 \\ 
         & 68 & 11.23 & 146.47 & 78.91 & 57.72 & 32.27 & 143.13 \\ 
         & 70 & 3.20 & 170.22 & 55.26 & 67.13 & 22.84 & 128.36 \\ 
         & 72 & 1.93 & 174.04 & 51.43 & 68.56 & 21.40 & 130.71 \\ 
         & 74 & 1.39 & 175.61 & 49.94 & 69.28 & 20.69 & 135.65 \\ 
\hline
        171.5 & 60 &  $>35$ &  & &  &  &  \\ 
         & 70 & 12.56 & 139.83 & 85.42 & 55.16 & 34.83 & 159.63 \\ 
         & 80 & 9.59 & 155.52 & 69.93 & 61.44 & 28.56 & 135.71 \\ 
         & 85 & 1.52 & 174.96 & 50.90 & 68.88 & 21.11 & 133.50 \\ 
         & 90 & 0.76 & 176.33 & 49.58 & 69.66 & 20.33 & 144.02 \\ 
\hline
175 &         75 & 22.42 & 136.93 & 88.05 & 54.31 & 35.66 & 159.79 \\ 
         & 80 & 10.79 & 143.92 & 81.31 & 56.91 & 33.08 & 153.15 \\ 
         & 85 & 8.70 & 154.62 & 70.83 & 60.95 & 29.04 & 139.61 \\ 
         & 90 & 1.96 & 173.26 & 52.55 & 68.18 & 21.81 & 133.75 \\ 
         & 95 & 1.01 & 175.98 & 49.87 & 69.42 & 20.57 & 139.86 \\ 
\hline
        180 & 80 & 19.98 & 137.16 & 87.82 & 54.26 & 35.71 & 161.78 \\ 
         & 85 & 10.82 & 144.43 & 80.79 & 57.18 & 32.80 & 147.58 \\ 
         & 90 & 9.25 & 145.65 & 79.67 & 57.71 & 32.28 & 151.31 \\ 
         & 95 & 7.47 & 156.48 & 69.11 & 61.64 & 28.35 & 135.49 \\ 
         & 100 & 1.79 & 173.47 & 52.42 & 68.26 & 21.74 & 135.25 \\
    \end{tabular}
\end{ruledtabular}
\end{table}

\FloatBarrier
\bibliography{VU_bibtex_master}

\end{document}